\documentclass[aps,twocolumn,prd,epsf]{revtex4}
\usepackage[latin1]{inputenc}
\usepackage[english]{babel}
\usepackage{latexsym}
\usepackage{amssymb}
\usepackage{graphicx}
\newcommand{\be}{\begin{equation}}
\newcommand{\ee}{\end{equation}}
\newcommand{\bea}{\begin{eqnarray}}
\newcommand{\eea}{\end{eqnarray}}

\begin{document}
%\begin{titlepage}
%\begin{titlepage}
%____________________________________________________________________________
\title{A model-independent dark energy reconstruction scheme using the
geometrical form of the luminosity-distance relation}
%____________________________________________________________________________
\author{Stéphane Fay$^{1,2}$\footnote{steph.fay@gmail.com} and
Reza Tavakol$^1$\footnote{r.tavakol@qmul.ac.uk}}
\affiliation{$^1$School of Mathematical Sciences,\\
Queen Mary, University of London,
London E1 4NS, UK}
\affiliation{$^2$Laboratoire Univers et Théories (LUTH), UMR 8102\\
Observatoire de Paris, F-92195 Meudon Cedex, France}
%____________________________________________________________________________
\begin{abstract}
We put forward a new model-independent reconstruction scheme
for dark energy which utilises the expected geometrical features of the
luminosity-distance relation. The important advantage of this scheme is that
it does not assume explicit ansatzes for cosmological
parameters but only some very general cosmological properties 
via the geometrical features of the reconstructed luminosity-distance relation. 
Using the recently released supernovae data by the Supernova Legacy Survey
together with a phase space representation,
we show that the reconstructed luminosity-distance curves 
best fitting the data correspond
to a slightly varying dark energy density with the Universe
expanding slightly slower than
the $\Lambda CDM$ model. However, the $\Lambda CDM$ model fits the data at
$1\sigma$ significance level
and the fact that our best fitting luminosity-distance curve is
lower than that of the corresponding $\Lambda CDM$ model could be due to
systematics.
The transition from an accelerating to a decelerating expansion occurs 
at a redshift larger than $z=0.35$.
Interpreting the dark energy as a minimally
coupled scalar field we also reconstruct the scalar field and its potential.
%Our reconstructions show 
%that any cosmological theories [DO WE WANT TO SAY THEORIES? YES] sharing 
%these properties [WHICH ONE? WE SHOULD SPECIFY "GEOMETRICAL PROPERTIES" SOMEWHERE] and fitting the SNLS data better
%than the $\Lambda CDM$ model cannot rule out this model
%at $1\sigma$ confidence level [REWRITE OK?].
We constrain $\Omega_{m_0}$ using the baryon acoustic
oscillation peak in the SDSS luminous red galaxy sample
and find that the best fit is obtained with $\Omega_{m_0}=0.27$,
in agreement with the CMB data.
\end{abstract}
\maketitle
%----------------------------------------------------%
\section{Introduction} \label{s0}
%----------------------------------------------------%
There is now overwhelming evidence from Supernovae data \cite{Rie04,Ast06}
as well as CMB measurements 
\cite{Spe03,Spe06}
and observations of large scale structure 
\cite{Tegmark-etal04,Seljak-etal05}
which suggest that the Universe is at present undergoing a phase 
of accelerated expansion. 
These data also provide evidence that nearly two thirds of the
total density of the Universe is in the form of an effective fluid with a 
negative pressure. 

Determining the underlying mechanism for this late--time                      
acceleration constitutes one of the fundamental                 
challenges facing cosmology today.
A large number of models have been proposed to account
for this acceleration. These fall into two main categories:
those that involve the introduction of an exotic matter source,
which include the cosmological constant and the quintessence 
scalar fields and those that involve changes to the gravitational 
sector of GR, either motivated by String/M theory, or through
ad hoc modifications of the Hilbert action.

Among these models some of the most commonly studied have been the so called
`dark energy' scenarios, which model the underlying accelerating agent as 
an effective perfect fluid with a negative equation of state (EOS).
In addition to their simplicity, one of the main reasons for their popularity
has been the fact that an effective model of this type can mimic 
a very wide variety 
of models, ranging from the cosmological constant with EOS equal to $-1$ to 
models with variable EOS such as quintessence models 
and brane inspired models such as that due to
Dvali-Gabadadze-Porrati \cite{DvaGabPor00,Fay06}.

Given the multiplicity of such model candidates, an urgent question 
at present is how to distinguish between them observationally.
Of particular interest is the nature of the EOS of
dark energy and importantly whether it is
variable and different from $-1$.

An important approach to this problem has been
to reconstruct the properties of the dark energy, including its
equation of state, from the Supernovae \cite{Ast06}, and more recently
the baryon acoustic oscillation (BAO) data \cite{Eis05}.
A large number of attempts have recently been made at such reconstructions. 
These include schemes that rely on specific functional ansatzes for 
the cosmological parameters to be reconstructed, such as for example 
the Hubble parameter \cite{state,SzyCza03} 
or the deceleration parameter \cite{ElgMul06}.
Other schemes employ more general parametrised forms such as interpolating 
fits with right behaviours at small and large redshifts \cite{Ala03,AlaSah04}.
There are also the so called model-independent schemes which attempt to recover 
the cosmological parameters directly from the data without assuming their forms
by using more general statistical tools, such as Markov Chain Monte Carlo techniques
\cite{WanMuk04} or Gaussian Kernels \cite{Sha05}.

In this paper we put forward a new model-independent reconstruction scheme 
which utilises the expected geometrical features of the luminosity-distance 
relation while fitting the observational data including 
the recently released Supernova Legacy Survey (SNLS) and 
baryon acoustic oscillation data.
Using a phase space representation
we reconstruct the luminosity-distance
curves best fitting the data and hence the 
corresponding cosmological parameters including
the Hubble parameter and the EOS.

The plan of the paper is as follows. In section \ref{s1} we describe 
our reconstruction scheme.
In Section \ref{s2} we give a brief account of the data sets used in our
reconstructions. Section \ref{s3} contains our results:
in subsection A, we present the results of 
our reconstructions of the cosmological 
parameters, assuming the cold dark matter density parameter to be 
$\Omega_{m_0}=0.27$. In subsection B we give a discussion of the 
degeneracy of the luminosity distance with respect to $\Omega_{m_0}$ 
and use the baryon acoustic oscillation data to constrain 
$\Omega_{m_0}$.
In subsection C we discuss the robustness of our 
results, and finally we conclude in section \ref{s4}.

%----------------------------------------------------%
\section{The reconstruction method}\label{s1}
%----------------------------------------------------%
A natural starting point in reconstruction schemes employing Supernova
observations is the luminosity-distance $d_l (z)$ expressed
in terms of the Hubble parameter $H$:
\be
\label{ld}
d_l (z)=c(1+z)\int^z_0\frac{dz}{H}\,.
\ee
Such reconstructions have taken a variety of forms, often relying on 
specific functional ansatzes for various cosmological parameters.
Here we propose an alternative method which instead of specifying 
precise functional ansatz for the luminosity-distance relation $d_l (z)$, 
imposes weak observationally motivated constraints on its geometry
given by:

\begin{enumerate}
\item[{\bf (a)}] $d_l'\ge 0$\,,
\item[{\bf (b)}] $d_l'' \ge 0$\,,
\item[{\bf (c)}]  $d_l''' \le 0$\,,
\end{enumerate}
where a prime denotes a derivative with respect to the redshift $z$.
These are rather natural and weak constraints motivated 
by current observations. In particular condition (a) is 
satisfied for any expanding Universe ($d_l'< 0$ 
always implies 
a contracting Universe). Condition (b) is 
satisfied by any Universe that is currently accelerating
(with a negative deceleration parameter $q<0$)
and which in the past tends to an Einstein-de Sitter model with $q=1/2$.
This can be seen by noting that using $dH/dt=-H^2(1+q)$,
the positivity of $d_l''$ implies 
$q<1$. Finally condition (c) is satisfied
by both Einstein-de Sitter and $\Lambda CDM$ models, for all redshifts $z$.
This can be seen by employing the statefinder parameter $r=\dddot a/(a H^3)$ \cite{state},
and using it to express the inequality (c) as $r\geq 3q^2+q-1$.
This inequality is satisfied, for all $z$, by the $\Lambda CDM$ model
(for which using the Hubble function $H^2=\rho_0 a^{-3}+\Lambda$ we have
$q=-1+24\rho_0e^{3\sqrt{\Lambda}(t+t_0)}/(e^{3\sqrt{\Lambda}(t+t_0)}+4\rho_0)^2$
and $r=1$, where $\rho_0$ and $t_0$ correspond to 
values at present time). The condition saturates for de Sitter space 
asymptotically and is satisfied by the Einstein-de Sitter model
(for which we have $q=1/2$ and $r=1$). In terms of the deceleration 
parameter, condition (c) imposes a lower bound $q'$ on $q$ such that 
$q'>\frac{q^2-1}{1+z}$.

Thus these conditions are compatible with 
current observations and are satisfied, for all $z$, by 
both Einstein-de Sitter and $\Lambda CDM$ models
which are the most commonly accepted models representing the early and
late dynamics of the Universe.
Together these features provide justification for their use in
constraining the reconstructed luminosity-distance curves,
which is the approach we shall take in the following.

Before we proceed with our reconstructions we 
recall that even though the reconstruction process itself
does not require a theoretical framework,
the physical interpretation of the 
reconstructed cosmological parameters
does so.
Given that the true underlying theory is not known
a priori, it is important that such a framework is general 
enough to accommodate a large enough range of 
possible candidates and yet at the same time 
is sufficiently simple to make it convenient to work with.
Here as our theoretical framework
we shall adopt general relativity plus a 
minimally coupled scalar field $\phi$ with a potential $V (\phi)$.
This simple framework includes an important set of candidates
including GR plus a cosmological constant and quintessence models.
For simplicity we shall consider a 
flat Friedmann-Lemaître-Robertson-Walker metric.
The dark energy can then be viewed as a perfect fluid
with the corresponding density, pressure and EOS given by
\be
\rho_\phi = \frac{1}{2}\dot\phi^2+ V, ~~ p_\phi =  \frac{1}{2}\dot\phi^2- V
, ~~ \mbox{w}_\phi = \frac{p_\phi}{\rho_\phi} \,, 
\ee
where a dot denotes a derivative with respect
to the proper time $t$.

To facilitate the interpretation of our results, we define 
the following well behaved expansion-normalised variables often used
in the dynamical systems analysis of such models 
\cite{copeland-etal,huey-tavakol}
\begin{equation}\label{vars}
u=\frac{\rho}{3H^2}, ~~
v=\frac{\sqrt{V}}{\sqrt{3}H}, ~~
w=\frac{\dot \phi}{\sqrt{6}H} \,,
\end{equation}
where $\rho$ is the matter (cold dark matter plus baryon) density.
In terms of these variables, the Friedmann and the $\dot H$
equations can be written as 
%------------------------------------%
\bea
\label{cons}
u&+&v^2+w^2=1\,, \\
\label{hd}
\frac{\dot H}{H^2}&=&\frac{3}{2}u+3v^2-3\,,
\eea
with the equation of state taking the form
\be
\label{eos}
\mbox{w}_\phi =\frac{(w^2-v^2)}{(w^2+v^2)}\,.
\ee
The reconstruction then proceeds by generating a 
large discrete set, $S_{dl}$, of luminosity-distance curves 
whose geometries are free apart from satisfying the weak 
constraints (a)-(c) given above as well as fitting the 
SNLS supernovae data set (see the next Section for the details). 
More precisely, each luminosity-distance curve $ d_l \in S_{dl}$
is constructed by a discrete set of points $P_i=(z_i,d_{l_i})$ 
satisfying the constraints (a)-(c). 
Let us assume for simplicity that $P_i$ are equally spaced points, such that $dz_i=z_{i+1}-z_{i}$ 
is a constant interval for all $i$. Then each successive point $P_i$ of
a luminosity-distance curve $d_l\in S_{dl}$ must satisfy the 
constraints (a)-(c), which when discretised are respectively 
given by
\bea
\label{discrete}
d_{l_{i+1}} &\ge & d_{l_i}  \nonumber \\
d_{l_{i+2}}-d_{l_{i+1}} &\ge& d_{l_{i+1}}-d_{l_i} \nonumber \\
d_{l_{i+3}} - 2d_{l_{i+2}}+d_{l_{i+1}} &\leq& d_{l_{i+2}} - 2d_{l_{i+1}}+d_{l_i}
\eea
Thus the reconstruction consists of finding all the luminosity-distance curves
defined by the discrete set of points $P_i$ respecting the above 
rules and fitting the supernovae data.
In practice we divided the interval $z\in [0,1]$ 
into steps of $dz_i=0.07$ for small $z$ and larger steps for larger $z$ since 
the slope of $d_l$ varies faster for the smaller $z$.
%in the first case with respect to the second one.

To reconstruct the discrete set of luminosity distance curves, we 
used an iterative scheme.
%a cellular automaton. 
The starting point is a luminosity distance 
curve defined by the points $P_i$ and located 
%"The starting point is a straight line curve defined by the points $P_i$ 
in the neighbourhood of and below the SNSL data. Then, 
following the above rules, the iterative scheme results in
% automaton constructs 
the next luminosity distance 
curve by moving up one of the points $P_i$. Hence, step by step, the slope of the 
reconstructed curves increase. Each curve thus defined is then
transformed, using an interpolation to go from the discrete 
form in terms of $P_i$ to a continuous one. We then check if the reconstructed curve
fits the data. 
If so, it is added to the set $S_{dl}$. The iterations stop when the slope of 
the reconstructed curve is sufficiently large to place it
everywhere above the data points, so that they are no longer fitted.
%until they fit [OK?] the SNLS data. 
%At that stage the reconstruction stops. Each curve thus defined is then
%transformed [OK?], using an interpolation to go from the discrete 
%form in terms of $P_i$ to a continuous one fitting [OK?] the SNLS data, 
%and then added to the set $S_{dl}$ [CHECK THIS REWRITE].
%[WE CAN POLISH THE ABOVE PARA TOGETHER]} 
Note that in the figures presented below, interpolation is used to 
give the reconstructed values at equally spaced steps of $dz_i=0.07$.
Once the set of curves $S_{dl}$ is determined, we calculate for each curve 
%------------------------------------%
\begin{equation}\label{hDl}
H=c\left[\left (d_l'-\frac{d_l}{1+z} \right )\frac{1}{1+z}\right]^{-1} \,,
\end{equation}
and
%------------------------------------%
\begin{equation}\label{hpDl}
H'=-c(1+z)\frac{2d_l+(1+z)\left[-2d_l'+(1+z)d_l''\right]}{(dl-(1+z)d_l')^2} \,.
\end{equation}
Expression (\ref{hDl}) together with (\ref{vars}) can then be used
to reconstruct $u(z)$ once $\Omega_{m_0}$ is specified.
Similarly the use of expression (\ref{hpDl})  together with 
the relation $d/dt=-(1+z)Hd/dz$ allows
$\frac{\dot H}{H^2}$ to be reconstructed. 
Equations (\ref{cons}) and (\ref{hd}) can then 
be used to reconstruct the variables $v$ and $w$ and 
hence from (\ref{eos}) the EOS $\mbox{w}_\phi$.
Thus the use of this scheme allows the cosmological parameters to be reconstructed,
without assuming specific ansatzes but only the well defined 
cosmological properties encoded in conditions (a)-(c).

A similar but technically different approach was used in
\cite{DalDjo05}. These authors reconstruct the
coordinate distance $y(z)$, which is related to luminosity-distance $d_l$ by the relation
$d_l=H_0^{-1}y(1+z)$, from supernovae and radio galaxy
data. In this scheme $y$ is fitted locally, over
redshift windows of typical length $\Delta z\approx 0.4$, by
using a quadratic fit
to reconstruct the curves $y(z)$ \cite{DalDjo03}.
The first and second derivatives of
$y$ are then used to reconstruct the potential and kinetic energy density.
As noted in \cite{DalDjo03}, a simpler approach would consist of
fitting a polynomial to the entire set of data. However, low order
polynomials would not be flexible enough whereas higher order polynomials
could induce unphysical oscillatory behaviours.
The main difference between our scheme and the one used by these authors
is that we fit the entire
data set by imposing a number of geometrical constraints on
the $d_l$ curves (and hence implicitly on $y$)
which is not necessarily representable by a
polynomial in $z$. This allows sufficient flexibility in order
to fit the entire data set while providing enough constraints to
eliminate unphysical oscillations of $d_l$.
As we shall see though similar our results do  not
totally agree. For example we obtain a much
weaker deviation from the $\Lambda$CDM model.

%----------------------------------------------------%
\section{Data}\label{s2}
%-----------------------------------------------------
In this section we briefly describe the cosmological data sets 
that were used in our reconstructions in the next section. 
%These include:
%\begin{itemize}
\begin{enumerate}

\item As our primary supernovae sample we took the first year data set 
from the Supernova Legacy Survey (SNLS) \cite{Ast06} 
with $71$ new supernovae below $z=1.01$, together with another $44$ 
low $z$ supernovae already available, i.e. a total of 115 SNe. 

\item To check the ability of our scheme to reconstruct data,
we employed a set of mock data. This consists of 115 supernovae 
with the same distribution in redshift and error bars as the SNLS data
but with the luminosity-distances replaced 
by those corresponding to the $\Lambda CDM$ model plus a gaussian noise.

\item To constrain $\Omega_{m_0}$ we also employed the baryon acoustic 
oscillation peak (BAO) 
recently detected in the correlation function of luminous red 
galaxies (LRG) in the Sloan Digital Sky Survey \cite{Eis05}. 
This peak corresponds to the first acoustic peak at recombination 
and is determined by the sound horizon. The observed scale of 
the peak effectively constrains the quantity 
$$
A_{0.35}=D_v(0.35)\frac{\sqrt{\Omega_{m_0}H_0^2}}{0.35c}=0.469\pm 0.017 \,,
$$
where $z=0.35$ is the typical LRG redshift 
and $D_v(z)$ is the comoving angular diameter distance defined as 
$$
D_v(z)=\left[D_M(z)^2\frac{cz}{H(z)}\right]^{1/3}\,,
$$
with 
$$
D_M=c\int^z_0\frac{dz}{H}\,.
$$
\end{enumerate}
To test the goodness of fit of our reconstructions we employ the standard
$\chi^2$ minimization. For the supernovae data, $\chi^2$ is defined by
$$
\chi^2=\sum_{i=1}^{n}\frac{(m^{obs}_i-m^{th}_i)^2}{\sigma^2_i}\,,
$$
where $n$ is the number of data, $m^{obs}$ and $m^{th}$ are respectively
the observed and the theoretically reconstructed
magnitudes and $\sigma_i$ is the uncertainty in the 
individual $m^{obs_i}$.
%----------------------------------------------------%
\section{Results}\label{s3}
%-----------------------------------------------------
To generate the set of luminosity-distance curves, $S_{dl}$,
we need to choose the number of points $P_i$
in order to describe the curves.
We found $7$ points to be sufficient for the results represented 
here. See, however, the subsection \ref{s33} below
for further discussion of this choice.

In our reconstructions, each luminosity-distance curve was fitted to 
the SNLS data.
Among the reconstructed curves $S_{dl}$ we only considered
those with $\chi^2<136.8$.
This upper bound on $\chi^2$ was chosen since for a model with 
one free parameter, such as a flat $\Lambda CDM$ model,
it corresponds to a reasonable value of $\chi^2$ per degree of 
freedom of $\chi^2_{DOF} =1.20$ \footnote{The number of degrees of freedom is defined as the 
number of data minus the number of free parameters in the model.
}. This would clearly be larger if $\chi^2>136.8$.

In the following we shall proceed in two steps. In subsection A
we shall assume $\Omega_{m_0}=0.27$ and reconstruct the cosmological
parameters using the weak constraints (a)-(c) given above.
In the subsequent subsection B we use the BAO data to constrain
$\Omega_{m_0}$ and show that $0.27$ is indeed the best fitting value.
Finally in subsection C we study the robustness of our results
with respect to the number of points used to define the
luminosity-distance curves, as well as the supernovae data set employed
and the precise value of $\Omega_{m_0}$ used.
%----------------------------------------------------%
\subsection{Reconstruction assuming $\Omega_{m_0}=0.27$}\label{s31}
%----------------------------------------------------%
The supernovae data alone do not constrain
the present values of the Hubble
function $H_0$ and its derivative $H_0'$.
This is due to the fact that
there is no data available, and hence no constraints, from the future. 
Consequently, if no further information is used,
the luminosity-distance curves can have any slope at present,
subject to the constraints (a)-(c).
Thus one needs to assume reasonable priors on $H_0$ and $H_0'$ 
when making reconstructions using only the supernovae data, 
without assuming additional information.

Here we shall make the following reasonable assumptions.
For the present value of the Hubble parameter we assume $60<H_0<80$, which
from (\ref{hDl}) implies that initially $3750<d_{l_0}'=\frac{c}{H_0}<5000$ Mpc. 
Since the derivative of the Hubble parameter is given by
$H_0'=-c(-2d_{l_0}'+d_{l_0}'')/d_{l_0}^{'2}$
and $d_{l_0}'$ can take very small values,
$H_0'$ can take large negative values at present.
From $H'H(1+z)=1/2\dot\phi^2+1/2\rho$, this would 
correspond to high negative values for the 
EOS \footnote{If $H'<0$, $\dot\phi^2<0$ and we have a ghost fluid}. 
Thus in order to have a reasonable lower bound on the present value of the EOS, 
we choose $H_0'>-40$, which from (\ref{hpDl}) implies 
$9375<d_{l_0}''=2d_{l_0}'-H_0'd_{l_0}^{'2}/c<13333$ Mpc or 
using (\ref{cons}-\ref{hd}) corresponds to $\mbox{w}_{\phi_0}>-2$
for the present value of the EOS.

\begin{figure}[h]
\centering
\includegraphics[width=6cm]{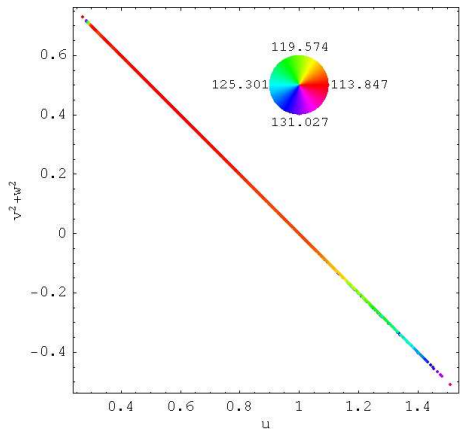}
\caption{\scriptsize{\label{const} 
Plot depicting $v^2+w^2$ versus $u$ for the reconstructed 
curves. The perfect straight line
provides a good confirmation for the constraint
$u+v^2+w^2=1$. 
The reconstructed points are shaded 
(coloured: references to colour refer to the web version with
colour figures)
according to their corresponding $\chi^2$ values,
with the deepest shade of grey (red),
representing the smallest values of $\chi^2$.}}
\end{figure}

The results presented below were obtained by 
reconstructing more than $23000$ luminosity-distance curves 
fitting the data. To show that all the reconstructed curves respect the 
Friedmann constraint (\ref{cons}), we have depicted in figure \ref{const} 
a plot of $v^2+w^2$ versus $u$ for these curves, which as can be seen
provides a very good confirmation of the constraint $u+v^2+w^2=1$.
Note that some reconstructions with negative energy density ($v^2+w^2<0$) are 
also in agreement with the data, but these correspond to
the larger values of $\chi^2$. This raises the interesting question 
of physicality of such solutions. We shall not consider 
this issue further here, but recall that negative energy densities 
sometimes arise, for instance in quantum cosmology or in 
considerations of braneworld models whose bulks
possess a negative cosmological constant.
%------------------------------------------------------------------------------
%------------------------------------------------------------------------------

\begin{figure}[h]
\centering
\includegraphics[width=6cm]{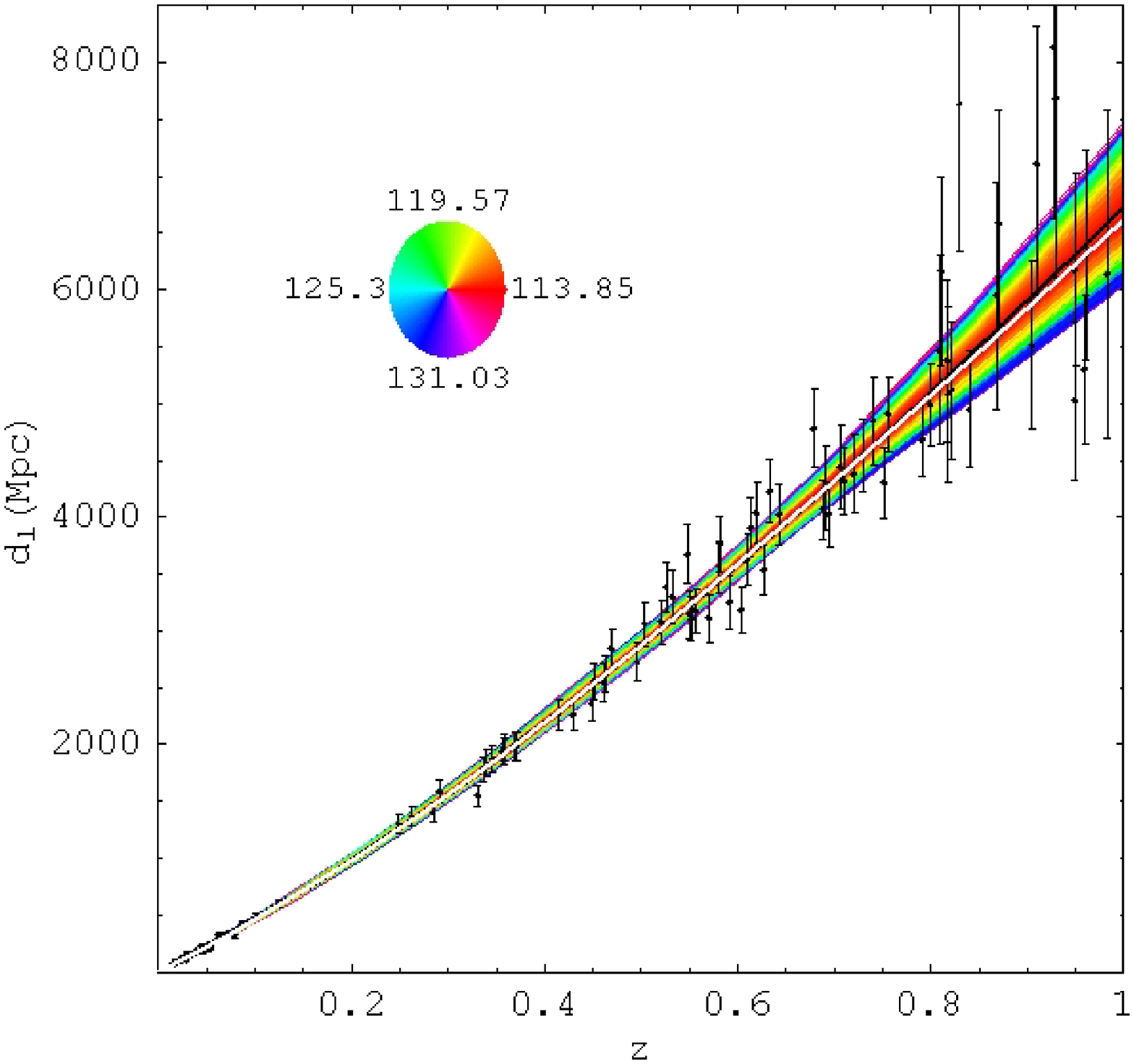}
\includegraphics[width=6cm]{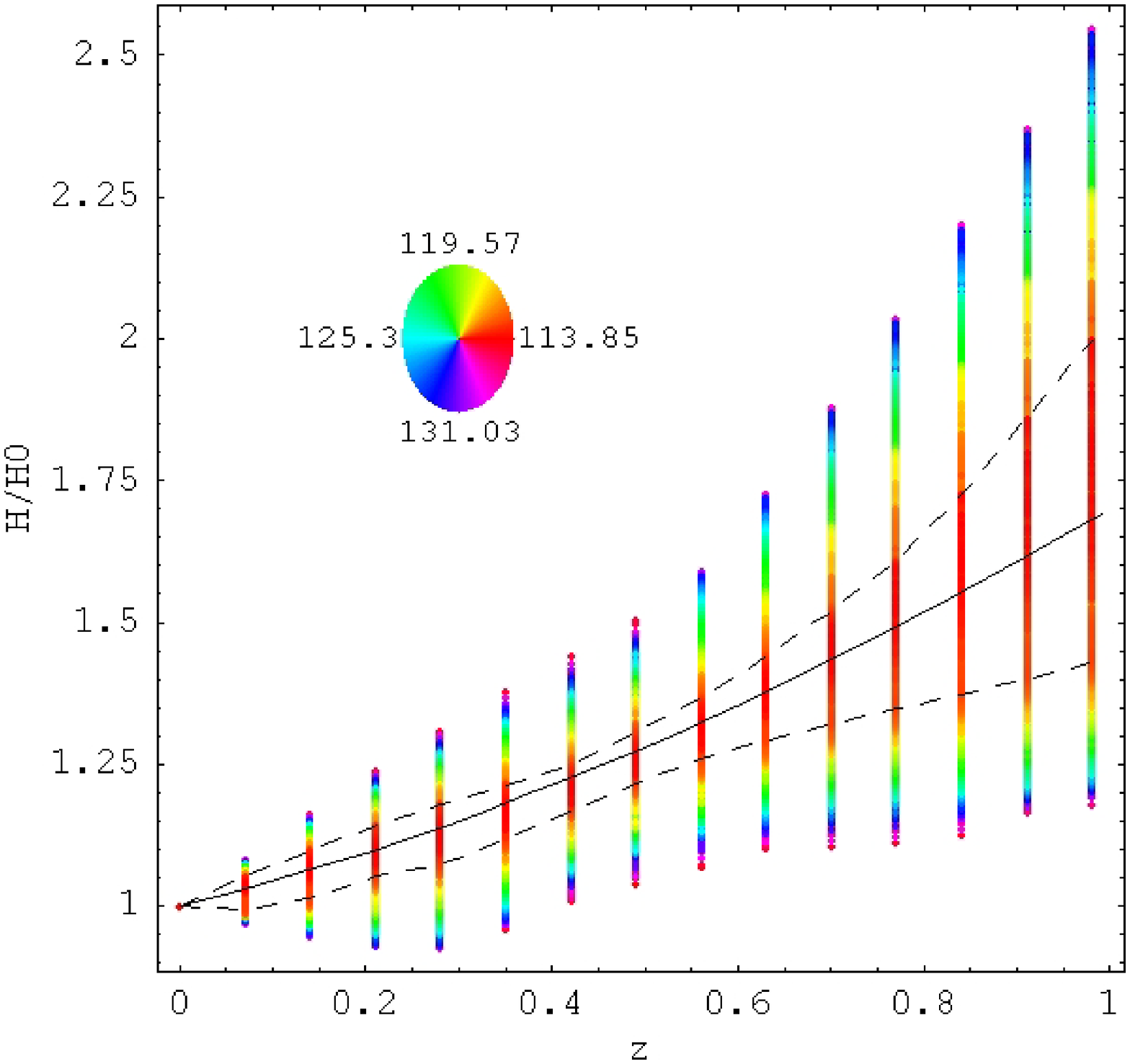}
\caption{\scriptsize{\label{dlH} 
Plots of the reconstructed luminosity-distance (top panel) and 
Hubble function (bottom panel) with respect to redshift. 
The dotted white (solid black) line 
represents the $\Lambda CDM$ model and the white line, 
the reconstructed luminosity-distance best fitting the data. 
For the Hubble function we have, for sake of clarity, 
plotted the points obtained numerically without joining them.}}
\end{figure}

\begin{figure}[h]
\centering
\includegraphics[width=6cm]{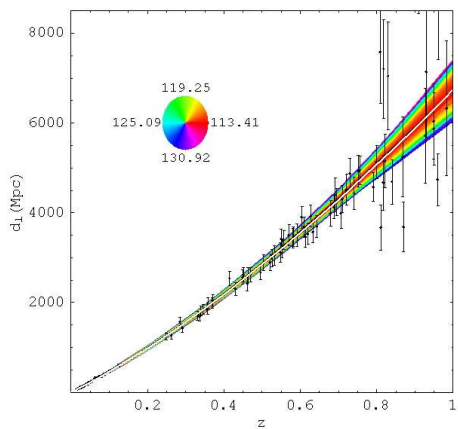}
\caption{\scriptsize{\label{dlmoch}
Plot depicting the reconstruction of the luminosity-distance relations
using the mock data. }}
\end{figure}

In figure \ref{dlH} we have plotted the reconstructed luminosity-distance
curves (top panel) and the corresponding Hubble parameters (bottom panel)
together with their $1 \sigma$ confidence levels (dashed lines)
respectively. 
As can be seen from the top panel, the $d_l$ curves 
best fitting the data have a slightly smaller $\chi^2$ 
(with the smallest having $\chi^2=113.85$) 
than the $\Lambda CDM$ model (with $\chi^2 \sim 114$).
%(which has $\chi^2_{DOF} \sim 1$). 
Hence, none of the reconstructed models 
rules out the $\Lambda CDM$ model at $1\sigma$.
The difference between our best fitting luminosity-distance 
curve and that of the $\Lambda CDM$ model becomes noticeable at large redshifts 
where the former curve is slightly lower. This deviation could come 
from systematics such as the Malmquist bias. 
Testing our reconstruction 
method with several samples
of mock data, we obtain curves which
very closely agree with the $\Lambda CDM$ model even at large redshifts.
However, they are very similar to the curves obtained using the real data.
This is due to the fact that the real data are themselves
very close to the $\Lambda CDM$ model, despite very small
deviations for large redshifts. An example of the luminosity-distance
reconstruction with mock data is displayed in figure \ref{dlmoch}.
We do not show the plots for the other physical quantities,
including the EOS, since they look very similar to the plots with
real data.

We also find the reconstructed Hubble parameters (bottom panel)
to have amplitudes which are in qualitative agreement with those 
found by other authors using different reconstruction 
techniques \cite{Ala03}. Here again, mock data produces similar results.
\\
%------------------------------------------------------------------------------
%------------------------------------------------------------------------------

%------------------------------------------------------------------------------
%---------------------------------------------------------------------------

\begin{figure}[h]
\centering
\includegraphics[width=6cm]{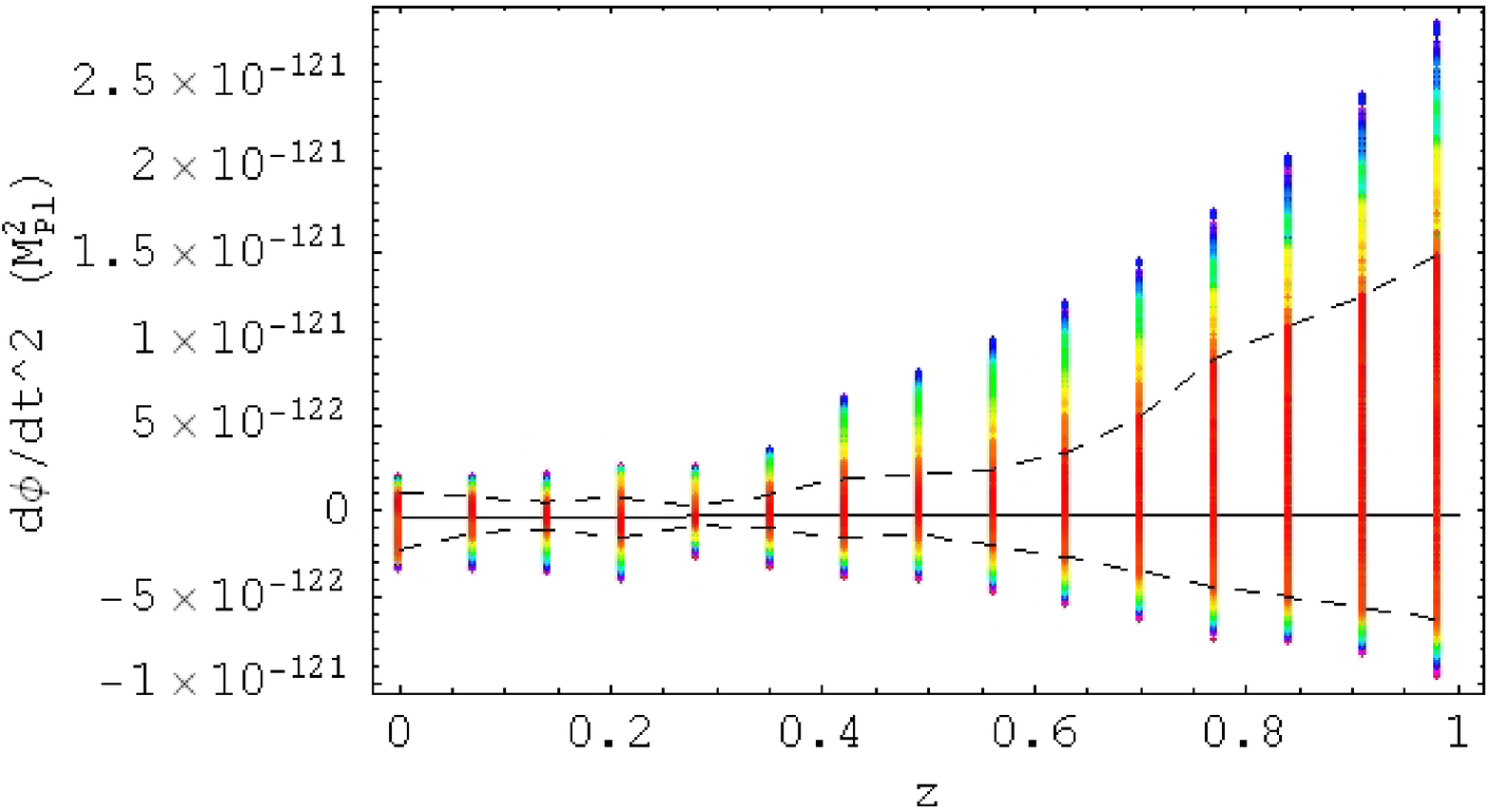}
\includegraphics[width=6cm]{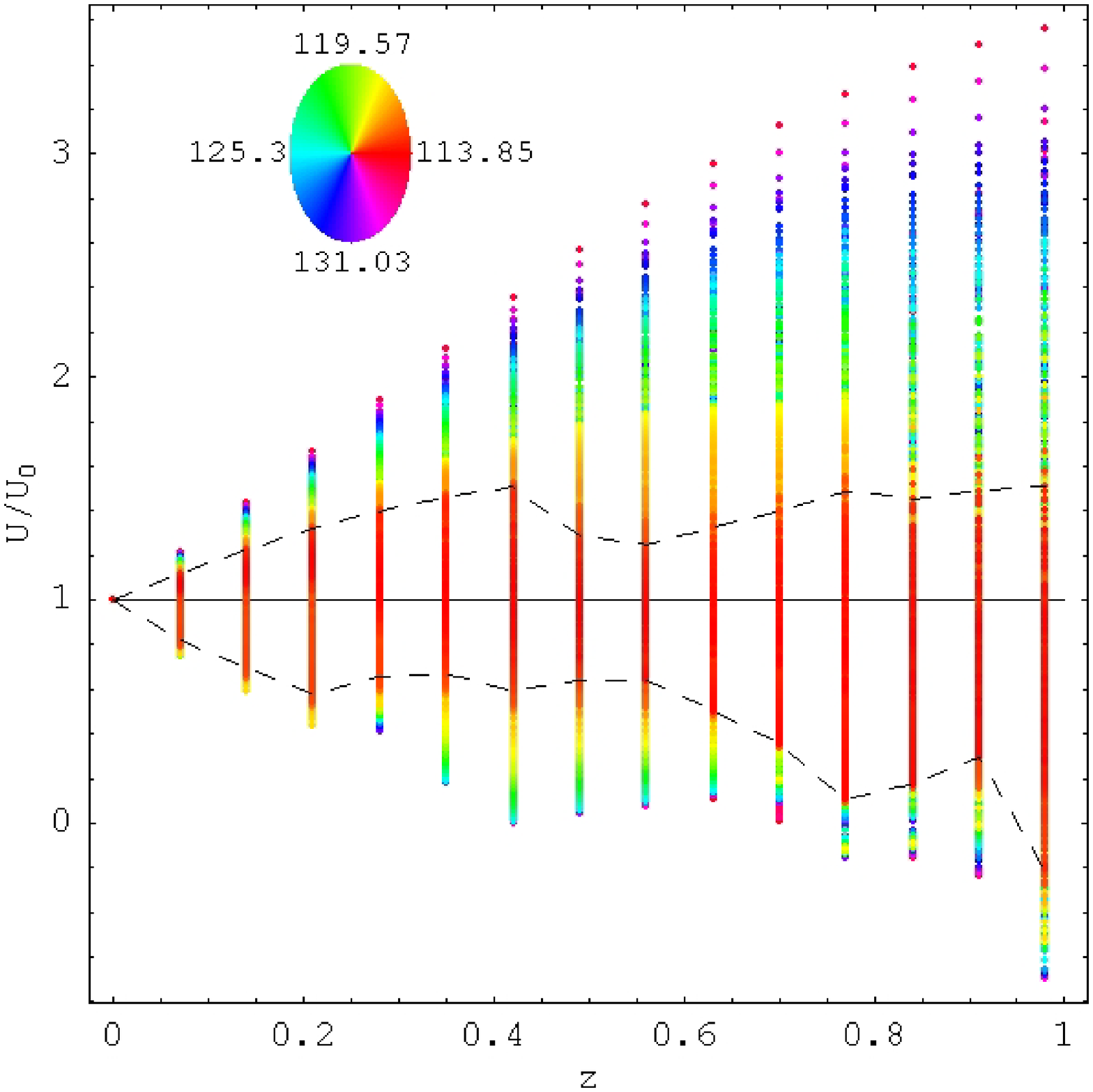}
\caption{\scriptsize{\label{Uphip2} Plots of all the reconstructed scalar 
field kinetic (top panel) and potential (bottom panel) energies
versus the redshift. The horizontal black lines represent the $\Lambda CDM$ model.}}
\end{figure}

\begin{figure}[h]
\centering
\includegraphics[width=6cm]{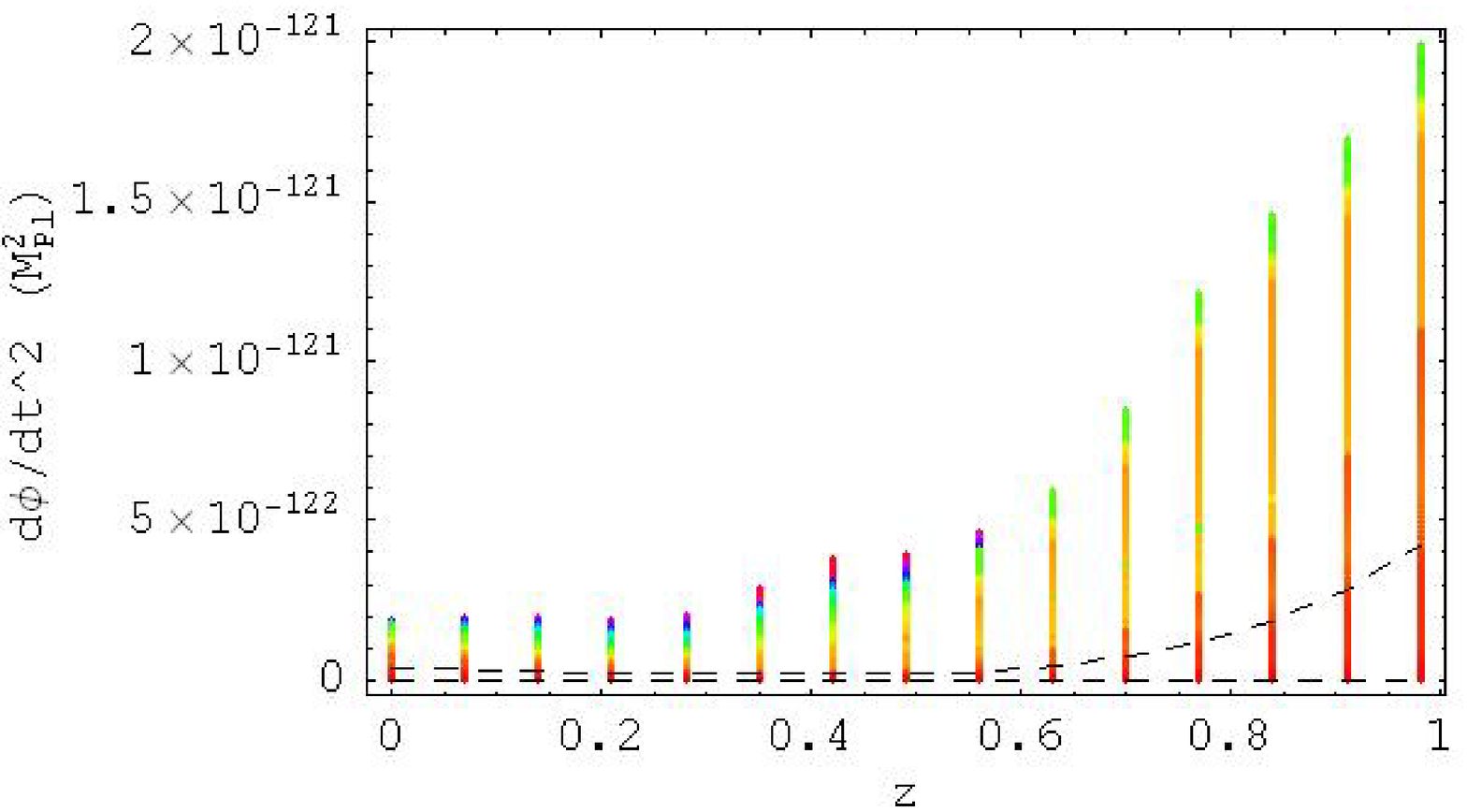}
\includegraphics[width=6cm]{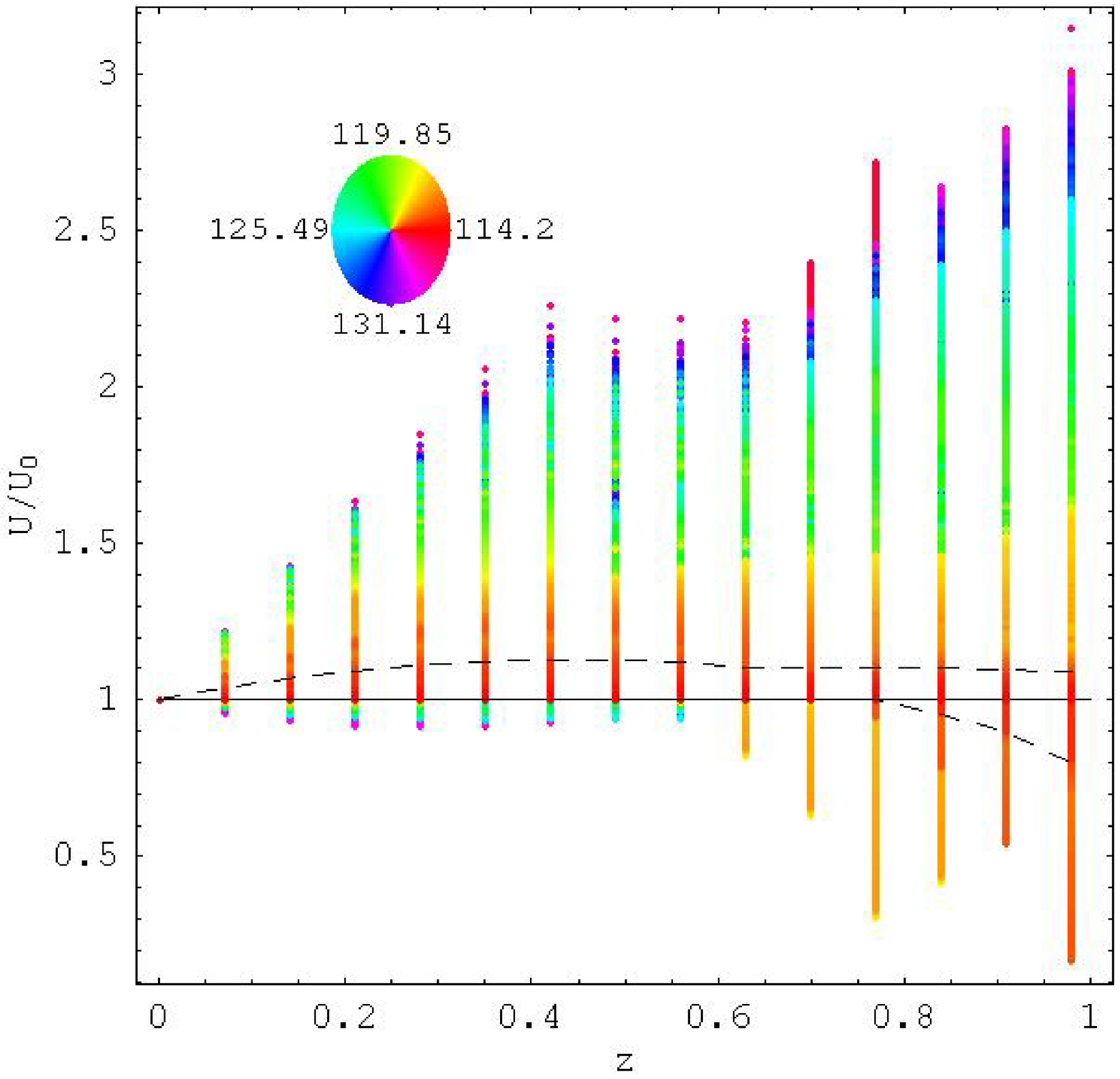}
\caption{\scriptsize{\label{Uphip2Phys} 
Plots of the reconstructed positive scalar field kinetic (top panel)
and positive potential (bottom panel) energies
versus the redshift. The horizontal black lines represent the $\Lambda CDM$ model.}}
\end{figure}

\begin{figure}[h]
\centering
\includegraphics[width=6cm]{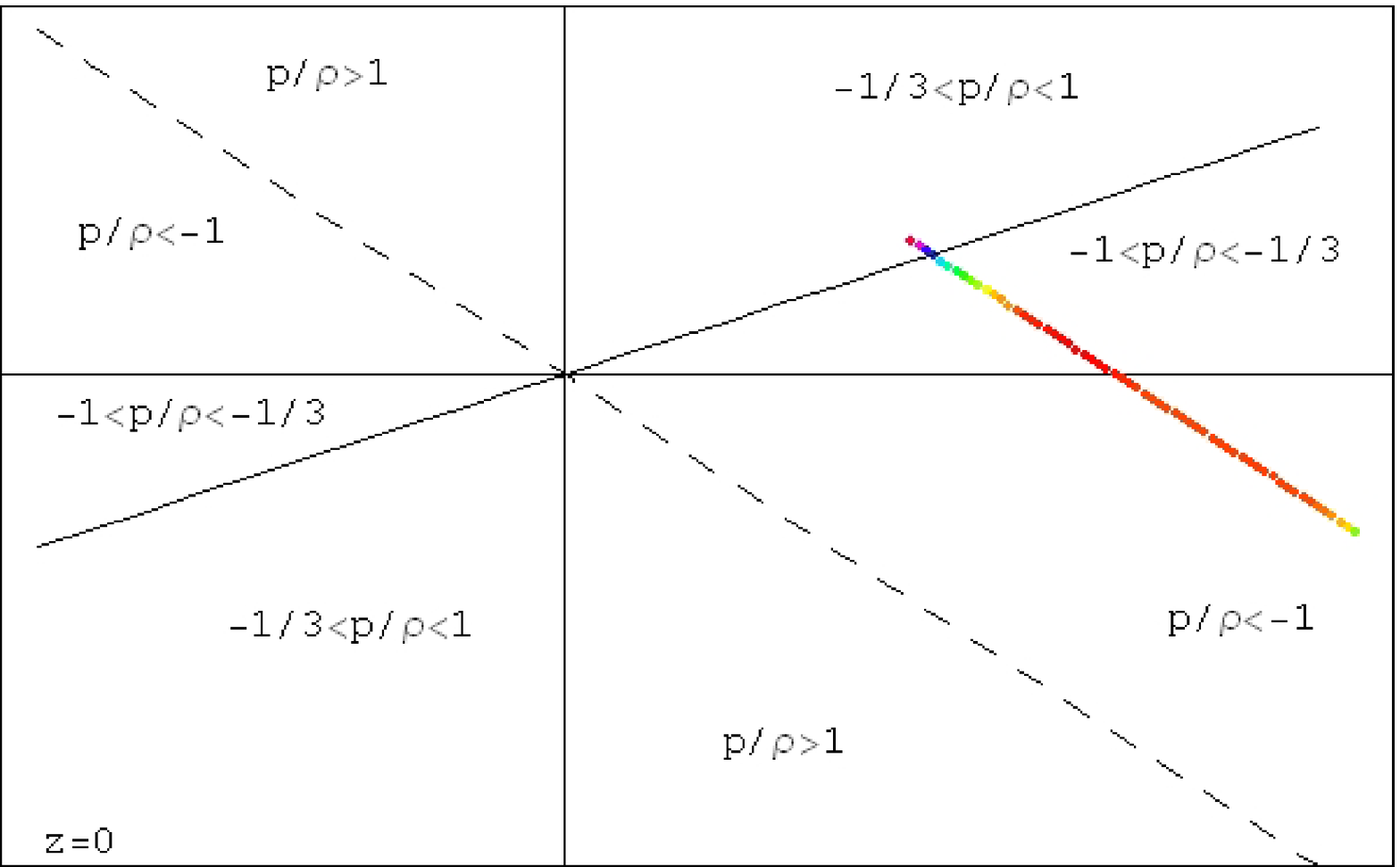}
\includegraphics[width=6cm]{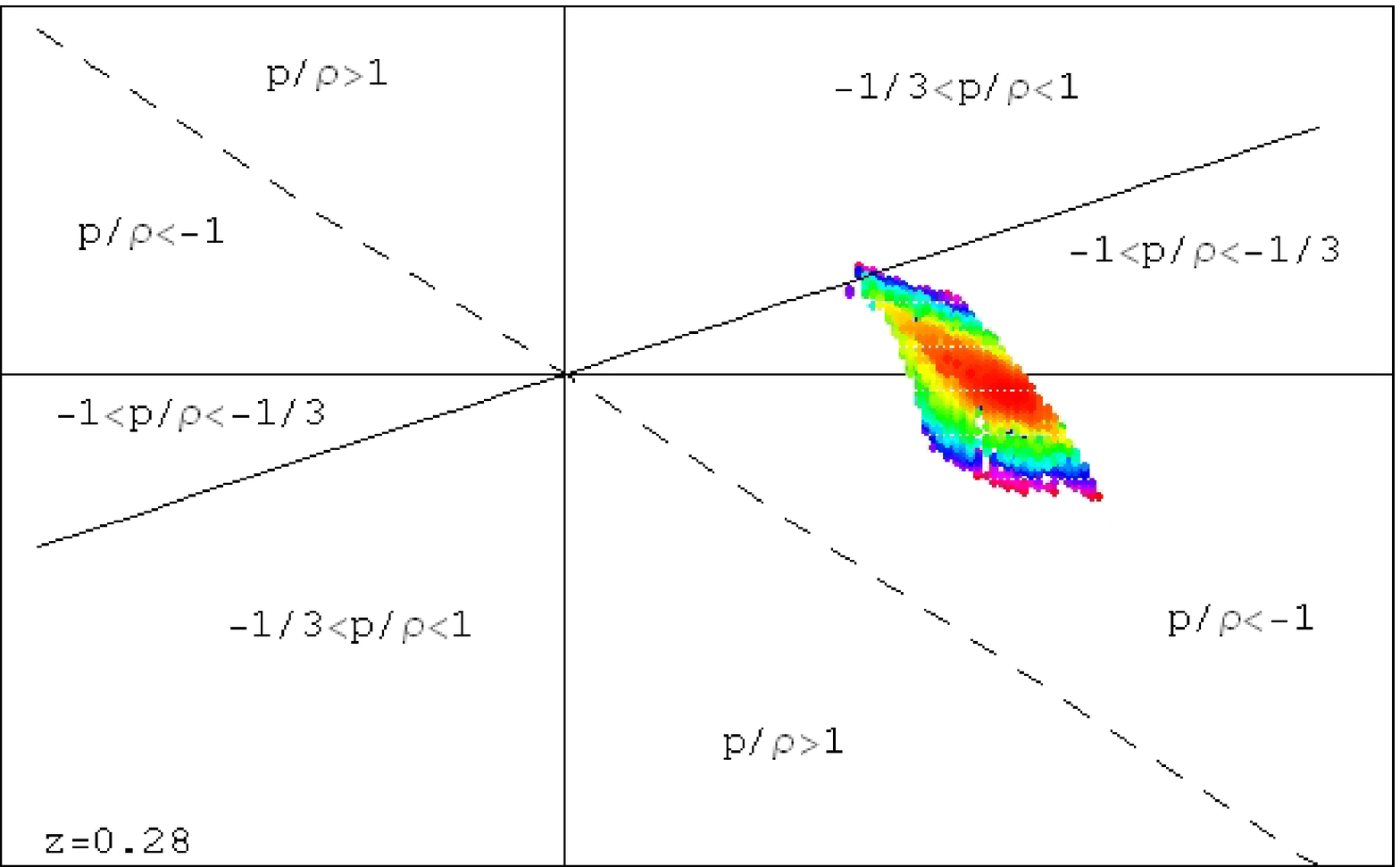}
\includegraphics[width=6cm]{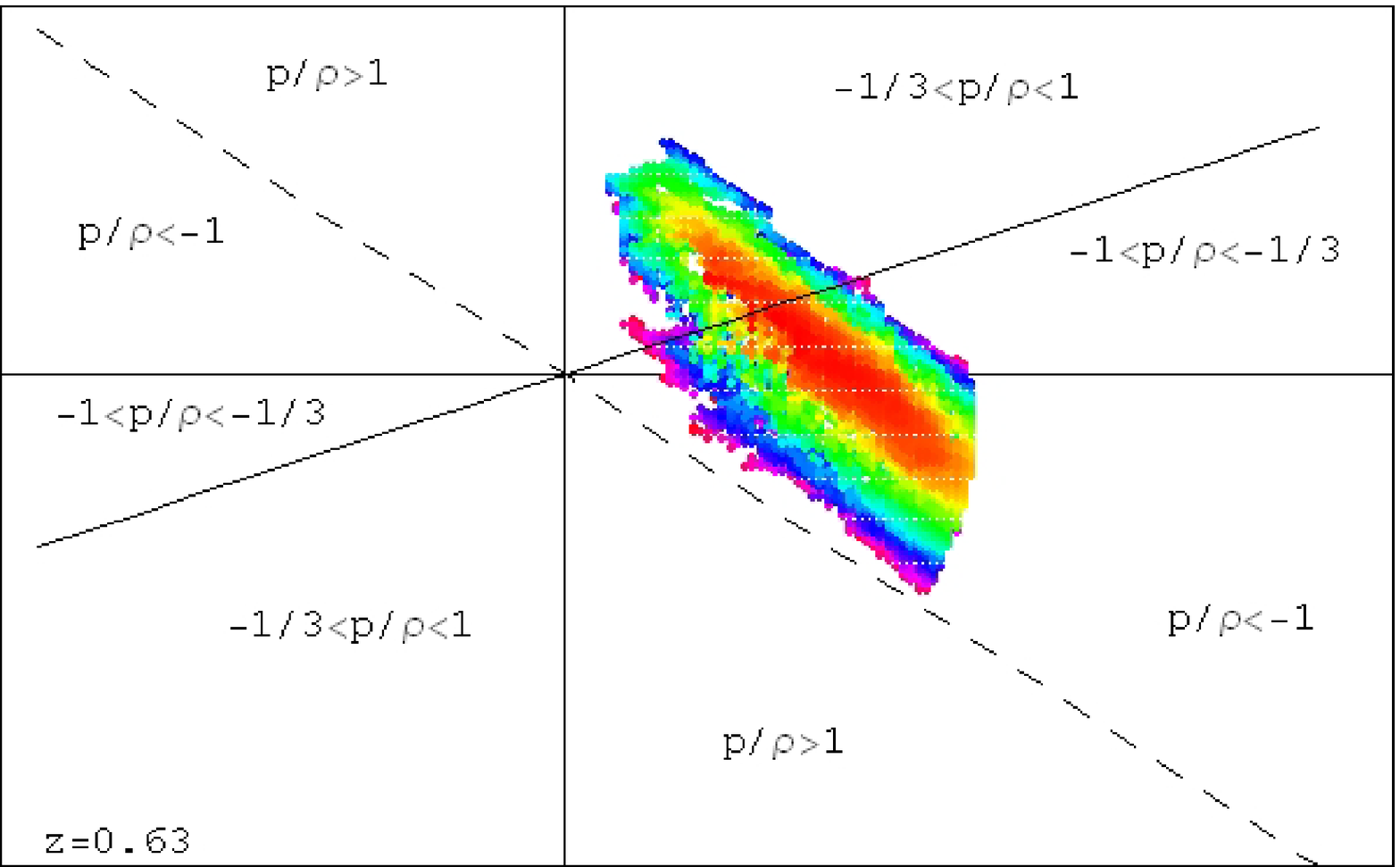}
\includegraphics[width=6cm]{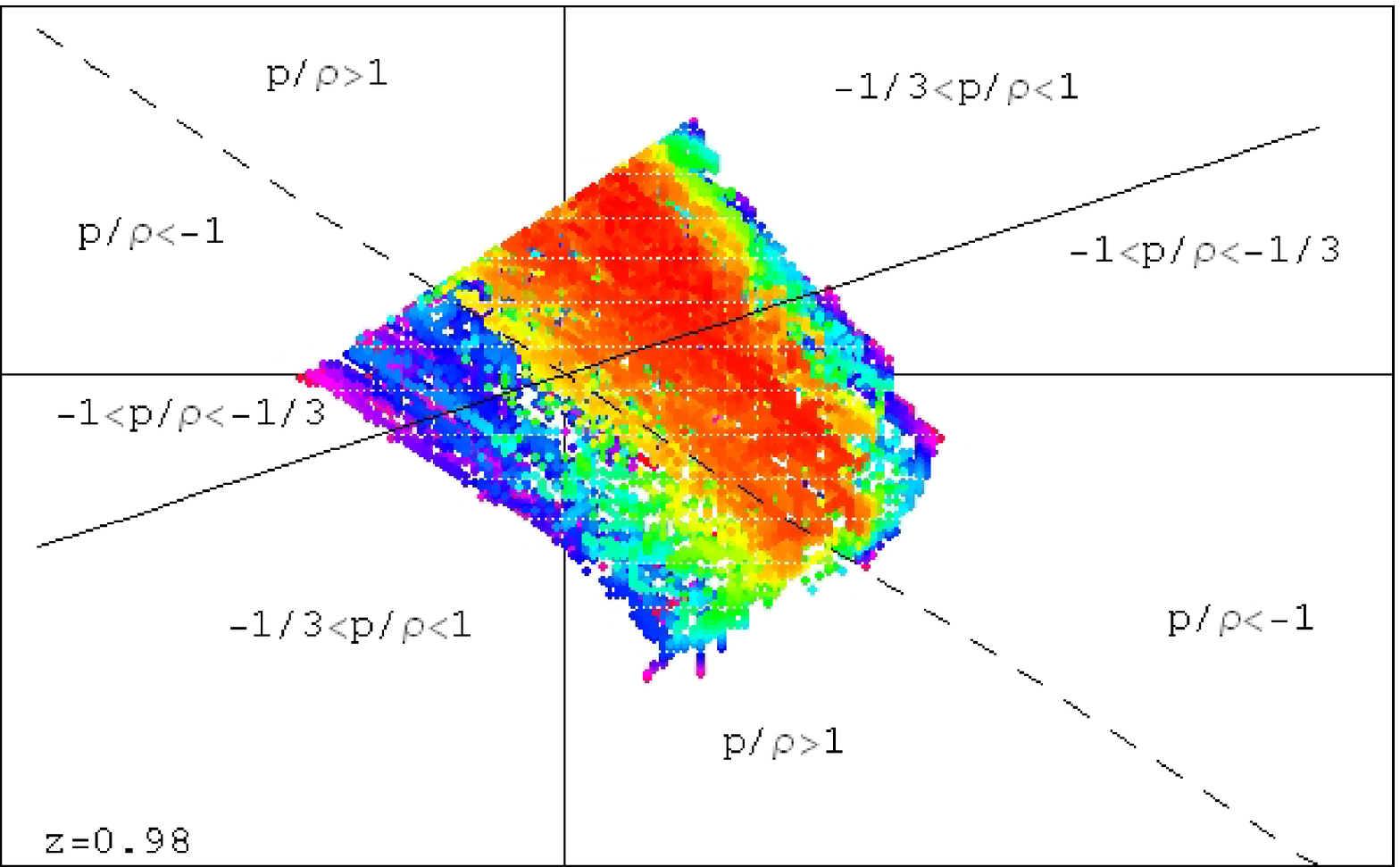}
\caption{\scriptsize{\label{y2z2Evol} Figures demonstrating the evolution of 
the reconstructed EOS (\ref{eos})
versus the redshift in the phase plane $(v^2,w^2)$}. The top panel,
corresponding to redshift $0$, shows the initial points which are forced into a linear
configuration due to the choice of $u_0=\Omega_{m_0}=0.27$ and the 
constraint equation $u+v^2+w^2=1$.
As the redshift increases (lower panels), the initial straight line 
configuration of points spreads
and eventually gets attracted to the neighbourhood of the line $w^2=v^2+1/3$
which corresponds to $d_l''=0$.}
\end{figure}

Taking as our theoretical framework GR plus a minimally
coupled scalar field, we also reconstructed the 
kinetic and potential energies corresponding to the 
scalar field $\phi$, together with their $1\sigma$ confidence 
levels \footnote{assuming a theory with one free parameter.}.
These are depicted in figure \ref{Uphip2} for the reconstructed $d_l$. 
However, to be able to interpret these reconstructions within
the framework of a minimally coupled scalar field, 
a number of physical constraints need to be taken into account \cite{Vik05},
such as those concerning the signs of the potential and kinetic terms. 
%$\dot \phi$ and $U$.
%In o must be taken into account in the 
%representation of $\dot \phi$ and $U$. 
To be cautions, we have extracted from the reconstructions
the curves corresponding to 
the quintessence scalar field models with a positive potential.
%These imply the removal of 
%reconstructions with negative potential and kinetic terms 
These are depicted in figure \ref{Uphip2Phys} showing that the best 
fitting curves, in the context of quintessence scalar field interpretation, 
are very close to the $\Lambda CDM$ model. This is particularly 
striking for $z<0.6$ where the degeneracy is very weak. 
For larger redshifts, some degeneracy appears due to larger 
error bars but remains relatively weak. 
\\
%-----------------------------------------------------------------------------
%-----------------------------------------------------------------------------
%-----------------------------------------------------------------------------
%-----------------------------------------------------------------------------
We also reconstructed the dark energy EOS 
(for {\it all} the reconstructed luminosity-distance curves)
whose evolution with redshift is depicted in figure \ref{y2z2Evol}.
To obtain a compact representation of the EOS,
we found it more convenient to have a $(v^2,w^2)$
plane representation instead of the usual plot of EOS as a function of $z$.
This choice of variables is well suited to the particular 
form (\ref{eos}) of the EOS in this case,
whose denominator can take small values
(for instance when the kinetic and potential terms
take similar values but with different signs) resulting
in large variations in the EOS as the redshift increases.
As can be seen from figure  \ref{y2z2Evol} this 
allows a compact representation of the EOS,
with the diverging values of EOS
corresponding to the dashed line.
In this figure the horizontal line ($w^2=0$) represents
the EOS for the cosmological constant ($\mbox{w}_\phi=-1$) and the
vertical line ($v^2=0$) the EOS for a
stiff fluid ($\mbox{w}_\phi=1$). The remaining line defines the
$\mbox{w}_\phi = -1/3$ line which
demarcates the limit between the accelerated and decelerated expansion.

The shaded (coloured) patches in the panels of figure \ref{y2z2Evol} represent, 
from top to bottom, the
reconstructed values of the EOS at increasing values of the redshift
given by $z=0, 0.28,0.63,0.98$.
Each point constituting
the shaded (coloured) patches corresponds to a reconstructed luminosity-distance curve,
with the deepest shades of grey (red) representing the best fits (lowest $\chi^2$ values).
The top panel depicts a cloud of initial points in the $(v^2,w^2)$ plane,
which since we are initially assuming $u_0=\Omega_{m_0}=0.27$, are forced
to lie on a straight line by the
Friedmann constraint $u+v^2+w^2=1$.
As we go to lower panels (higher redshifts), $\Omega_{m}$ evolves
differently for each point, resulting in the dispersal of the initial
straight line into different clouds of points. 
Now since we are assuming $d_l'''\le 0$,
$d_l''$ becomes smaller as the redshift increases, 
but always stays positive.
As a result the initial straight line configuration of points spreads
and eventually gets attracted to the neighbourhood of the line $w^2=v^2+1/3$
which corresponds to $d_l''=0$. To see this,
recall that using (\ref{hd}) and (\ref{ld})
we have $\frac{H}{c}d_l''=2+\dot HH^{-2}$. This together with (\ref{cons}) and
(\ref{hpDl}) then shows that $d_l''=0$ implies $w^2=v^2+1/3$.
As can be seen from the bottom panel,
this line acts as the accumulation end state
of the initial cloud of points.

To summarise, the evolution of EOS with $z$ depicted in these panels demonstrates
that as $z$ increases, the points corresponding to the $d_l$ best fitting
the data stay well centred around the cosmological constant
line ($w^2=0$) in agreement with the behaviour found for the 
scalar field kinetic and potential energies.

\begin{figure}[h]
\centering
\includegraphics[width=8cm]{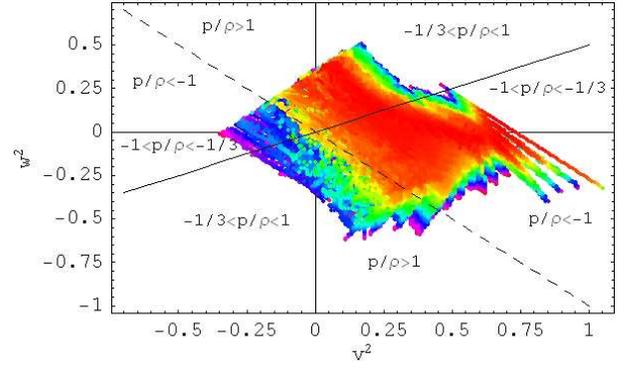}
\caption{\scriptsize{\label{y2z2} Figure showing the superposition of
similar panels to figure \ref{y2z2Evol} for all intermediate values of $z \in [0, 1]$
in steps of $dz=0.07$. As can be seen the distribution
of the points with the deepest shade of grey (red),
corresponding to the best fits,
is well centred around the cosmological constant line.}}
\end{figure}

We have also depicted in figure \ref{y2z2} the superposition of 
panels similar to those depicted in figure \ref{y2z2Evol} 
for intermediate values of 
the redshift considered in the range $z \in [0, 1]$, in steps of $z=0.07$.
We note that some possible variation of dark energy has been reported 
using reconstruction techniques 
different from that employed in this paper \cite{WanFre04,Ala03, AlaSah04}. 
Such variations are also possible according to our results
at $1\sigma$ confidence level, but are severely constrained 
in the context of quintessence scalar field models,
as can be seen from Fig. \ref{Uphip2Phys}.
%--------------------------------------------------------------------------
%--------------------------------------------------------------------------
%--------------------------------------------------------------------------
%--------------------------------------------------------------------------

\begin{figure}[h]
\centering
\includegraphics[width=6cm]{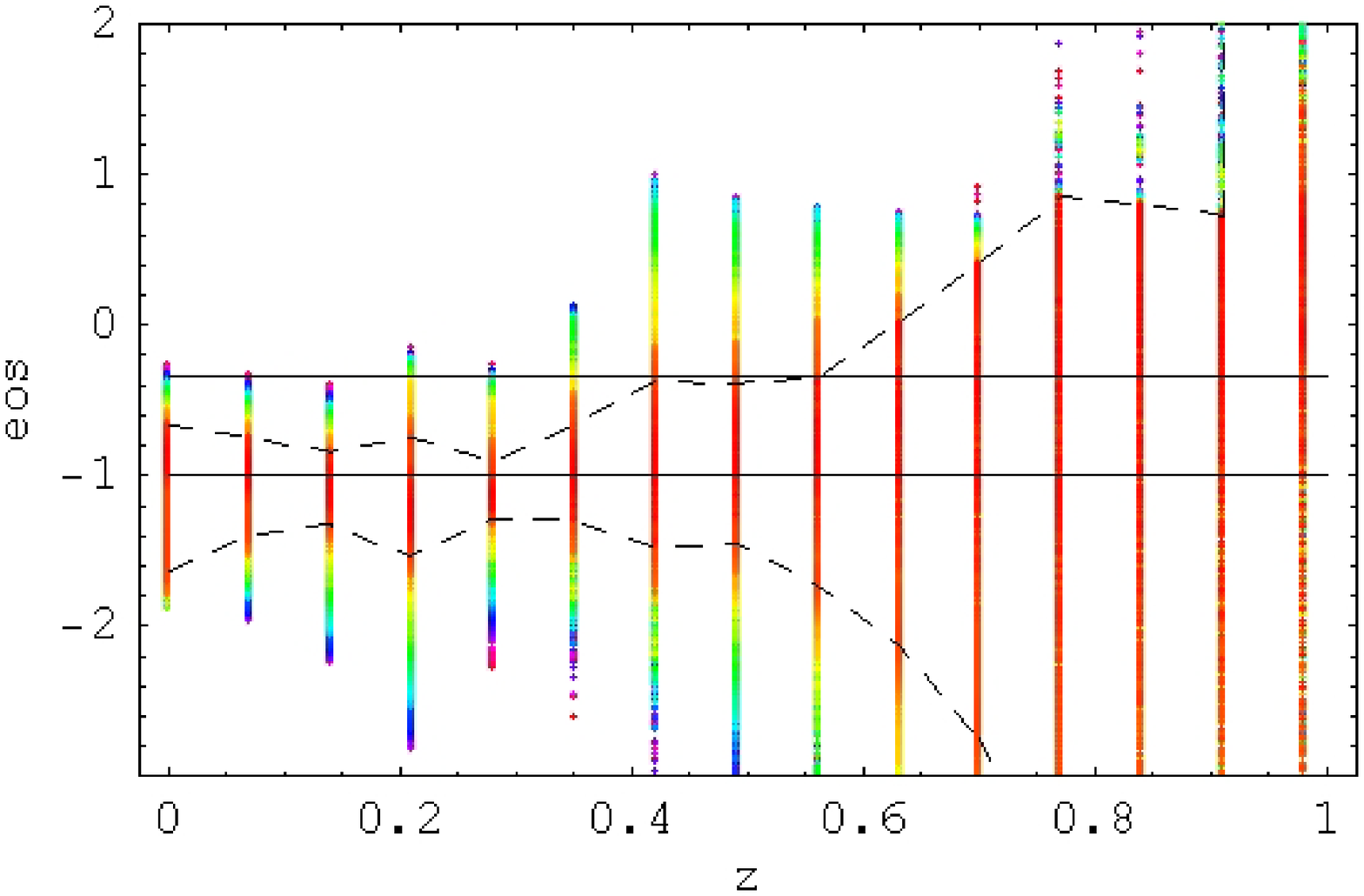}
\includegraphics[width=6cm]{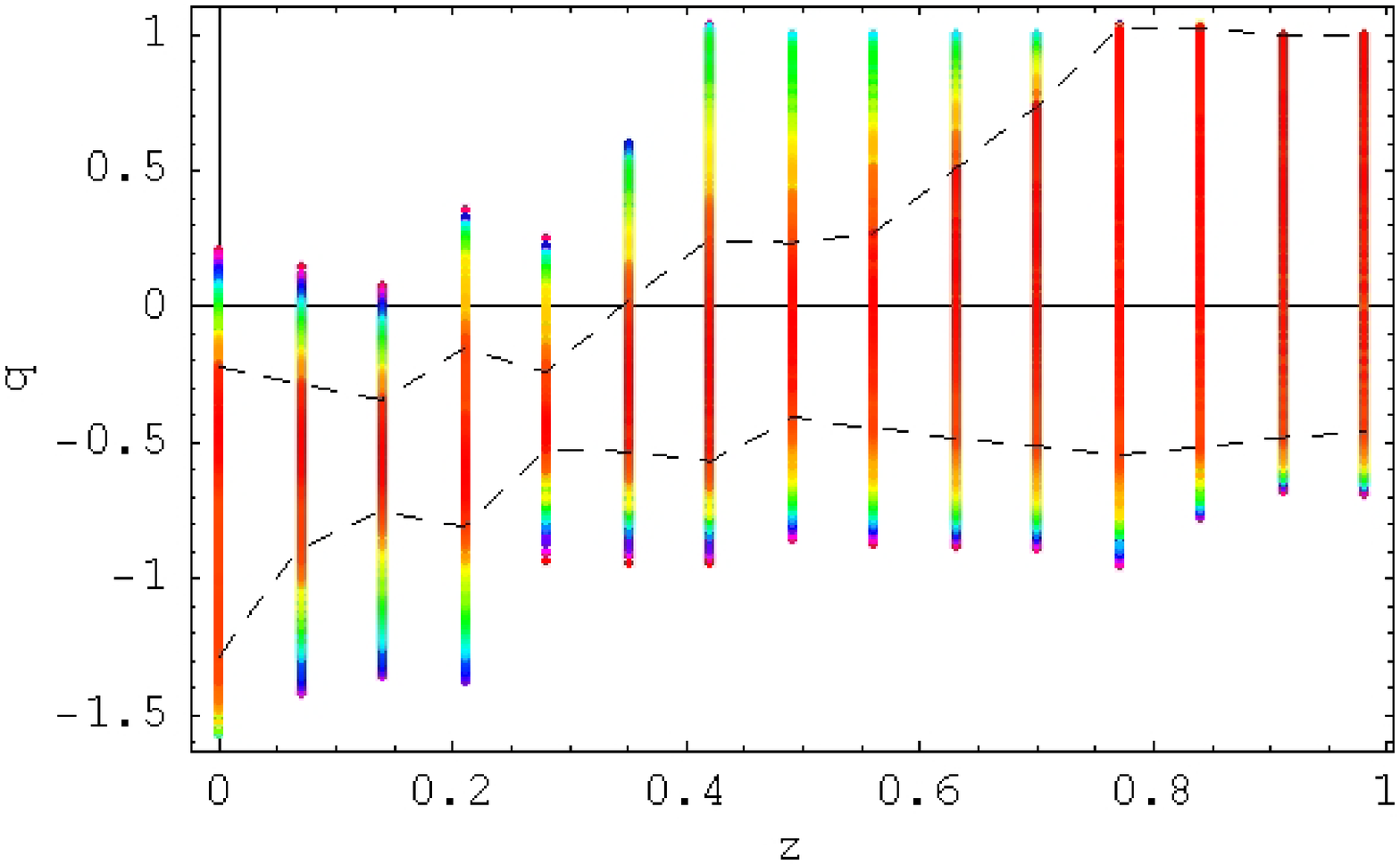}
\caption{\scriptsize{\label{bestEos}The evolutions of the reconstructed EOS corresponding to the 
luminosity-distance curves best fitting the data as a function of 
the redshift (top panel) and
the corresponding deceleration parameter (bottom panel) with their 
$1\sigma$ confidence levels (dashed lines).}}
\end{figure}

\begin{figure}[h]
\centering
\includegraphics[width=6cm]{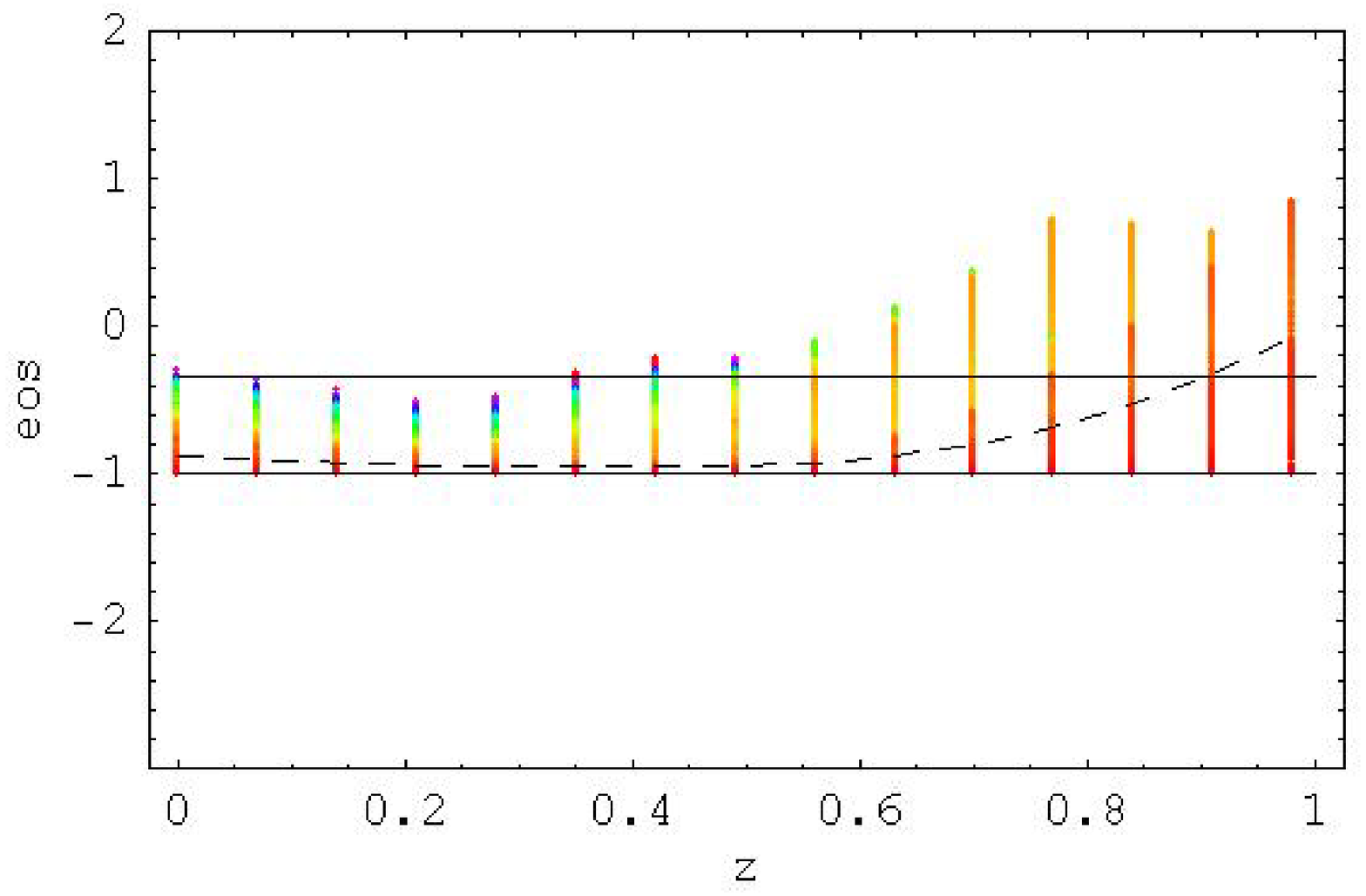}
\includegraphics[width=6cm]{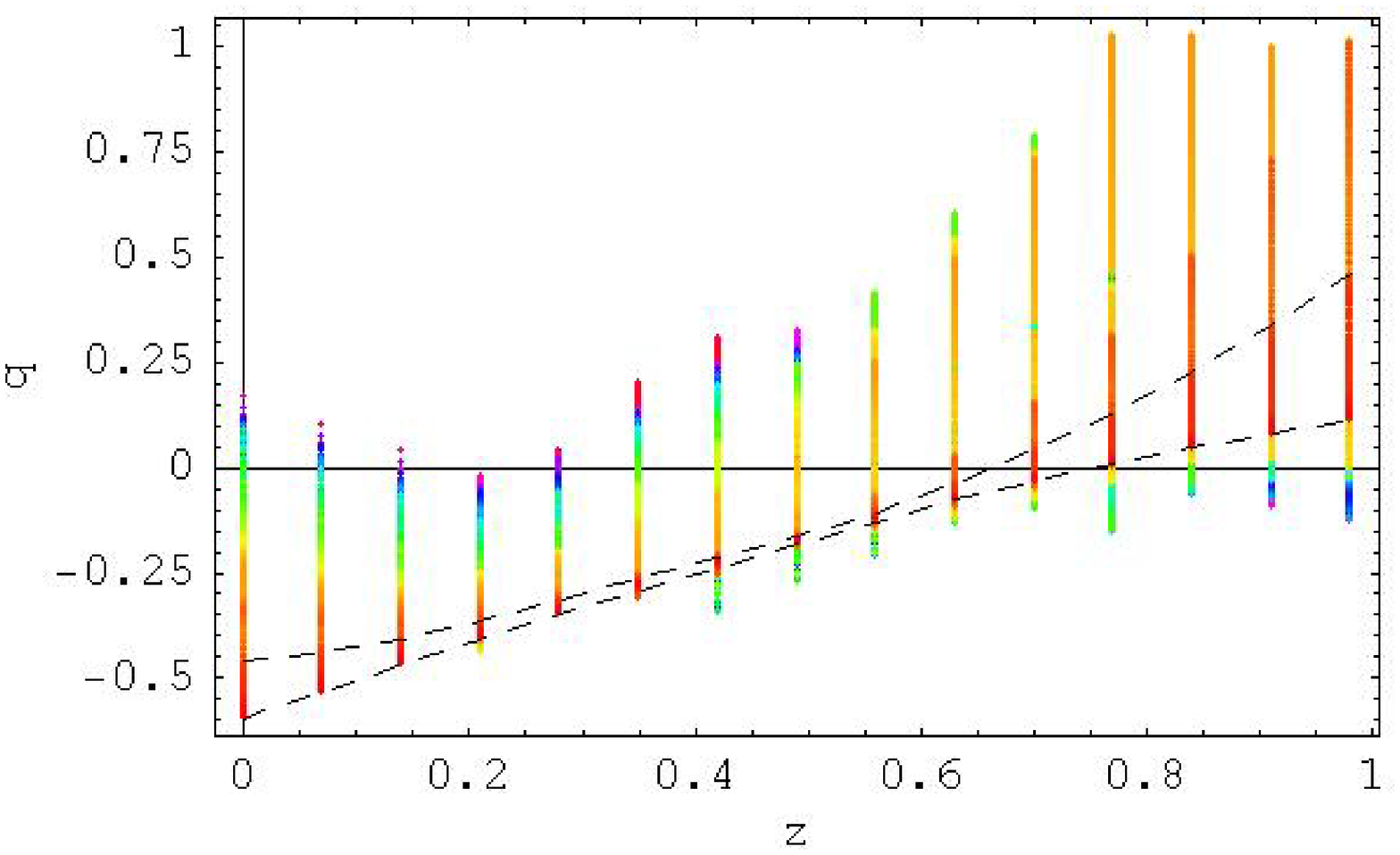}
\caption{\scriptsize{\label{bestEosPhys}
The evolutions of the reconstructed EOS corresponding to the 
luminosity-distance curves best fitting the data as a function of 
the redshift (top panel) and
the corresponding deceleration parameter (bottom panel) with their 
$1\sigma$ confidence levels (dashed lines), using the scalar field representation 
confined to quintessence models.}}
\end{figure}

We have also plotted in (top panel of) figure \ref{bestEos} the best 
fitting reconstructions for the EOS as a function of redshift with the $1\sigma$ confidence level.
As can be seen, initially ($z\sim 0$) the EOS takes values 
around $\sim -1$. Then, around the redshift of $0.45$, 
the degeneracy increases becoming very large beyond $z=0.8$. 
This follows the evolution of error bars and noise levels in the data. 
We have also plotted the evolution of the associated deceleration parameter 
(depicted in the bottom panel of the figure \ref{bestEos}). 
This shows that a transition from an accelerated to a decelerated expansion
would occur after the redshift of $0.35$. There can, nevertheless,
exist solutions which do not undergo such a transition, 
even at $1\sigma$ confidence level, in agreement with \cite{ElgMul06}.
This shows that constant as well
as negative deceleration parameters are also compatible with
the supernovae data, thus indicating that at present the
supernova data does not establish with certainty a transition
from accelerated to decelerated expansion.\\
%--------------------------------------------------------------------------
The panels in figure \ref{bestEosPhys}
also show the reconstructed EOS and the associated deceleration parameter 
when the reconstructions were confined
to the quintessence scalar field models with 
positive potentials. 
As can be seen at $1\sigma$ confidence level, 
the reconstructed EOS is very close to that of the 
$\Lambda CDM$ model but could vary weakly beyond $z\simeq 0.6$.
Moreover the deceleration can begin for a redshift $z > 0.65$.
%--------------------------------------------------------------------------
%----------------------------------------------------%
\subsection{Constraining $\Omega_{m_0}$}\label{s32}
%----------------------------------------------------%
Our reconstructions in the previous subsection did not provide
any information about the $CDM$ density parameter $\Omega_{m_0}$ 
which was assumed to be $0.27$. This is due to the fact that the 
luminosity-distance relation is determined purely by the 
Friedmann equation which is highly
degenerate with respect to this parameter.
To see this, let us consider a dark energy model
with a constant EOS, $\mbox{w}_\phi= \Gamma-1$, and the 
corresponding Friedmann equation
$$
(H/H_0)^2=\Omega_{m_0}(1+z)^3+\Omega_{\phi_0}(1+z)^{3\Gamma}
$$
Re-writing the $CDM$ density parameter as 
$\Omega_{m_0} = \Omega_{1m_0}+\Omega_{2m_0}$, the Friedmann equation becomes
$$
(H/H_0)^2=\Omega_{1m_0}(1+z)^3+\Omega_{2m_0}(1+z)^3+\Omega_{\phi_0}(1+z)^{3\Gamma}
$$
This form of $H(z)$ may be viewed as a new dark energy
model with a different $CDM$ density parameter $\Omega_{1m_0}$,
and a different dark energy density $\rho_\phi$
represented by the last two terms in this expression.
Hence without changing the Hubble function we can re-group the 
last two terms as a new dark energy term thus:
$$
(H/H_0)^2=\Omega_{1m_0}(1+z)^3+\Omega_{1\phi_0}\frac{\rho_\phi}{\rho_{\phi_0}}
$$
with a corresponding equation of state given by
$$
p_\phi/\rho_\phi=\frac{\Omega_{2m_0}(1+z)^3+\Gamma\Omega_{\phi_0}(1+z)^{3\Gamma}}{\Omega_{2m_0}(1+z)^3+\Omega_{\phi_0}(1+z)^{3\Gamma}}-1
$$
Thus given a Hubble function we can construct different 
representations with different effective $CDM$ density parameters
and dark energy components, but with identical
luminosity-distance functions $d_l$. This makes transparent 
the fact that the luminosity-distance $d_l$ is potentially
highly degenerate with respect to the $CDM$ density in Universe. 
It also illustrates the fact that a constant EOS can misleadingly 
become time-dependent if the matter density is incorrectly 
chosen, as is shown in \cite{Sha05}. We should, however, add
that since in practice $\Omega_{m_0}$ cannot be determined 
precisely, this demonstrates that a real observer can never prove that 
$\mbox{w}_\phi$ is truly constant, similar to the 
limitations that exist in determining the exact flatness of the  
Universe from observations. In practice one can only take the 
most likely value for $\Omega_{m_0}$
given by the most accurate available
observations in order to determine the corresponding
EOS.
 
\begin{figure}[h]
\centering
\includegraphics[width=8cm]{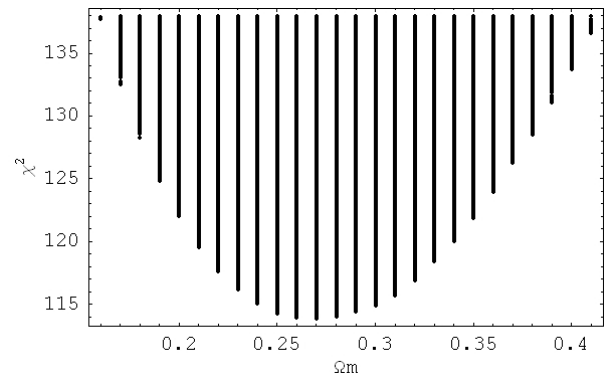}
\caption{\scriptsize{\label{omega} 
Figure depicting the variation of reconstructed $\Omega_{m_0}$ 
with respect to $\chi^2$,
using the SNLS and BAO data.}}
\end{figure}

To constrain $\Omega_{m_0}$, we need further input 
from observations. Here we used the recent baryon oscillation data \cite{Eis05}
(discussed in Section III), to further constrain
each luminosity-distance curve. 
We found that for our reconstructed set of curves $S_{dl}$, 
$\Omega_{m_0}$ lies in the range $0.16<\Omega_{m_0}<0.41$, with the best 
fit obtained for $\Omega_{m_0}=0.27$. This can be seen clearly from the plot
of $\Omega_{m_0}$ versus $\chi^2$ shown in 
figure \ref{omega} which justifies the choice of the specific value $\Omega_{m_0}=0.27$ 
in the previous subsection.
%----------------------------------------------------%
%----------------------------------------------------%
%----------------------------------------------------%
\subsection{Robustness of the reconstruction}\label{s33}
%----------------------------------------------------%
In this subsection we study the robustness of our results obtained
in the previous subsections.

In our reconstructions so far we have used $7$ points ($d_l,z$)
to describe the luminosity-distance curves. 
This choice was made by trial and error. 
We find that by taking too many points we can "over-fit" the data,
in the sense that we can always obtain a perfect but artificial 
match to any set of data. In that case one would in effect be
fitting noise with an anomalously low best 
$\chi^2$ which can, among other things, rule out 
the $\Lambda CDM$ model at $1\sigma$.
On the other hand, taking too few points would amount to "under-fitting" the data.
Thus one would not be able to fit the model 
represented by the data and the best $\chi^2$ can be anomalously high. 
For example, it could make it impossible to fit the $\Lambda CDM$ model.
In order to determine the optimal number of points, the use of 
mock data is extremely useful. We have tested our reconstruction scheme
with several sets of mock data, all based on the $\Lambda CDM$ model. 
Using seven points we were able to recover, in all cases,
the $\Lambda CDM$ model at $1\sigma$ with a best $\chi^2$ (typically around $113.7$)
only a few tenths of percent smaller than $114$ (which is close to the corresponding value 
for the $\Lambda CDM$ model). This demonstrates that, given the 
distribution of the SNLS data and the corresponding error bars,
our reconstruction process works well 
(although a slight over-fitting is unavoidable in practice as 
one is not expected to recover the best fit value exactly).
We checked that our results remain robust
with small changes in the number of points.

To summarise, care must be taken in choosing the number of
points taken to define a luminosity-distance curve. We found that
the employment of the mock data together with our conditions (a)-(c)
allow an appropriate number of points to be chosen, i.e. around $7$.

We note that the number of points defining a luminosity-distance curve is
not related to the degrees of freedom of the underlying theory giving 
rise to that curve. Thus, 
a straight line can be defined by $7$ points even though only $2$ points
are necessary to construct the line, while 
analytically the equation of a straight line
passing through the origin only has one
degree of freedom. 

% We should note also that $7$ points would lead to overfitting 
%in cases where the error bars are larger than those of the SNLS sample. 
%For instance, we have checked that this is the case with Riess data \cite{Rie04}. 
%In that case the signal of the underlying theory fitting the data is always nearly
%covered by the noise [WE NEED TO SAY WHAT NEEDS TO BE DONE THERE? 
%I THINK WE SHOULD DROP RIESS DATA TO CLARIFY THE DISCUSSION WHICH IS 
%ALREADY DIFFICULT TO FOLLOW]. 

We also checked that reasonable changes in the value of $\Omega_{m_0}$ does not alter
the qualitative behaviour of dark energy EOS found here.
%----------------------------------------------------%
\section{Conclusion}\label{s4}
%----------------------------------------------------%

We have proposed a model-independent reconstruction scheme which
is compatible with current observations and shares
a number of geometrical features with the Einstein-de Sitter
and $\Lambda CDM$ models, which are the most commonly accepted
models representing the early and late dynamics of the Universe.
Together these features provide justification for our 
proposed scheme.

Using this scheme, and assuming $\Omega_{m_0}=0.27$ together with the 
SNLS supernovae data, we have reconstructed a large set of 
luminosity-distance curves. Using these curves we
have reconstructed the cosmological parameters, including
the EOS. Our reconstructions show that the luminosity-distance curves 
best fitting the data correspond 
to a slightly varying dark energy density with the Universe 
expanding slightly slower than
the $\Lambda CDM$ model. However, the $\Lambda CDM$ model fits the data at 
$1\sigma$ and the fact that our best fitting luminosity-distance curve is
lower than that of the corresponding $\Lambda CDM$ model could be due to 
systematics such as Malmquist bias. Reconstructing the EOS, 
large degeneracy appears around $z=0.6$ in agreement with increasing 
error bars and noise levels in the distribution of data. The reconstructed 
deceleration parameter shows that the transition redshift to a 
decelerating Universe should be larger than $z=0.35$.

Assuming the theoretical framework to be GR plus a minimally coupled scalar 
field, we also considered reconstructions confined to quintessence models 
with positive potentials.
%also considered the case with positive values of the
%reconstructed kinetic and potential terms.
%and potentials of the scalar field\cite{Vik05}. 
In that case we find the best fitting reconstructed curves to be 
very close to those corresponding to the $\Lambda CDM$ model, 
in particular for low redshifts $z<0.6$. For larger redshifts, 
larger error bars in the data allow a small increase in the 
kinetic term and EOS and a small decrease in the potential term at $1\sigma$. 
However, in the context of quintessence interpretation, 
it seems rather hard to find a model fitting the data better than the 
$\Lambda CDM$ model at $1\sigma$.

%We have checked that our results are robust with respect
%to over-fitting or under-fitting phenomena discussed above. 
%They also remain qualitatively unchanged with reasonable changes
%in the value of $\Omega_{m_0}$ assumed.

We also used the BAO data to constrain $\Omega_{m_0}$ 
and found that $\Omega_{m_0}$ takes values
in the range $0.16<\Omega_{m_0}<0.41$ with a best fit
given by $\Omega_{m_0}=0.27$, in close agreement with CMB data \cite{Spe03,Spe06}.

Using different techniques, and employing supernovae as well as
other data, other authors \cite{AlaSah04,Ala03,WanFre04,WanMuk04} also find evidence for 
a possible variation of dark energy density. However, a constant dark energy 
density is again never ruled out to a
satisfactory confidence level and thus the important question of whether the
EOS varies in time or not still remains open.

The present paper takes a new step in reconstructing the equation of state by
providing a new model-independent reconstruction scheme
of the cosmological parameters. 
Our approach can be applied to other theoretical
frameworks which can be cast into GR plus a perfect fluid.
It would be interesting to repeat these reconstructions with 
the next set of SNLS data to determine 
whether the best fitting curves move closer or
further away from the $\Lambda CDM$ model.
%----------------------------------------------------%
\section*{Acknowledgments}
%----------------------------------------------------%
We would like to thank B. Bassett, R. Daly,
G. Djorgovski, S. Gilmour, R. Maartens, L. Perivolaropoulos,
D. Polarski, V. Sahni and A. Shafieloo for helpful comments.
S. F. is supported by a Marie Curie Intra-European Fellowship of the 
European Union (contract number MEIF-CT-2005-515028). 

%----------------------------------------------------%
\bibliographystyle{unsrt}

\end{document}